\documentclass[aps,prl,twocolumn,superscriptaddress,superscriptreference]{revtex4-2}
\usepackage{amsmath,bbold,bm,amssymb,scalerel,mathtools}
\usepackage{graphicx}
\usepackage{color}
\usepackage{enumitem}
\usepackage{algorithm,algpseudocode}
\usepackage{multirow}
\usepackage{colortbl,booktabs}
\usepackage{placeins}
\usepackage[usenames,dvipsnames]{xcolor}
\usepackage{tikz}

\usepackage{pdfpages}
\usepackage{pgffor}
\makeatletter
\AtBeginDocument{\let\LS@rot\@undefined}
\makeatother

\usepackage[colorlinks, linkcolor=blue!50!black, urlcolor=blue!50!black, citecolor=blue!50!black]{hyperref}
\tikzset{%
	every neuron/.style={
		circle,
		draw,
		minimum size=0.5cm
	},
	neuron missing/.style={
		draw=none, 
		scale=1.5,
		text height=0.333cm,
		execute at begin node=\color{black}$\vdots$
	},
}

\newcommand\equalhat{%
	\let\savearraystretch\arraystretch
	\renewcommand\arraystretch{0.3}
	\begin{array}{c}
		\stretchto{
			\scalerel*[\widthof{=}]{\wedge}
			{\rule{1ex}{3ex}}%
		}{0.5ex}\\ 
		=%
	\end{array}
	\let\arraystretch\savearraystretch
}

\usepackage[normalem]{ulem}

\begin{document}

\title{Unsupervised machine learning for supercooled liquids}
\author{Yunrui Qiu}
\affiliation{Department of Chemistry, Theoretical Chemistry Institute, University of Wisconsin-Madison, Madison, Wisconsin 53706, United States}

\author{Inhyuk Jang}
\affiliation{Department of Chemistry, Theoretical Chemistry Institute, University of Wisconsin-Madison, Madison, Wisconsin 53706, United States}

\author{Xuhui Huang\footnote{Corresponding author:xhuang@chem.wisc.edu}}
\affiliation{Department of Chemistry, Theoretical Chemistry Institute, University of Wisconsin-Madison, Madison, Wisconsin 53706, United States}
\affiliation{Data Science Institute, University of Wisconsin-Madison, Madison, Wisconsin 53706, United States}

\author{Arun Yethiraj\footnote{Corresponding author:yethiraj@wisc.edu}}
\affiliation{Department of Chemistry, Theoretical Chemistry Institute, University of Wisconsin-Madison, Madison, Wisconsin 53706, United States}

\date{\today}

\begin{abstract}
Unraveling the relation between structural information and the dynamic properties of supercooled liquids is one of the grand challenges of physics.  Dynamic heterogeneity, characterized by the propensity of particles, is often used as a proxy for the dynamic slowing down.
In this work, we introduce an unsupervised machine learning approach based on a time-lagged autoencoder (TAE) to elucidate the effect of structural features on the long-time dynamic heterogeneity of supercooled liquids. The TAE uses an autoencoder to reconstruct features at time $t + \Delta t$ from input features at time $t$ for individual particles, and the resulting latent space variables are considered as order parameters. In the Kob-Andersen system, with a $\Delta t$ about a thousand times smaller than the relaxation time, the TAE order parameter exhibits a remarkable correlation with the long-time propensity.  We find that radial features on all length-scales are required to capture the long-time dynamics, consistent with recent simulations. This shows that fluctuations of structural features contain sufficient information about the long-time dynamic heterogeneity.
\end{abstract}

\maketitle

The glass transition is a fascinating phenomenon in physics.  As a liquid is cooled, for some substances, crystallization is avoided and an amorphous solid state is reached.  The static structure, for example, the powder x-ray diffraction pattern, is similar to that of a liquid, but the dynamic properties, for example, the viscosity, are slower by several orders of magnitude.  Understanding the mechanism of the glass transition is one of the grand challenges in liquid state physics.  

Supercooling of the liquid is accompanied by the emergence of dynamic heterogeneity: molecules in some regions exhibit active re-arrangement, while molecules in other regions are almost frozen on the time-scale of the experiment.\cite{ediger2000spatially}   
It is often suggested that the weak correlation between traditional measures of ``structure" and ``dynamics" can be attributed to this dynamic heterogeneity.  In recent years, significant effort has been dedicated to elucidating the correlation between dynamical heterogeneity and structural properties.\cite{tanaka2019revealing}   In this work, we introduce an unsupervised machine learning method to estimate the long time dynamic heterogeneity from short simulations of the liquid.

An important advance in addressing the structural origin of dynamic heterogeneity is the concept of propensity of motion\cite{tanaka2019revealing,widmer2008irreversible} .  The propensity is obtained from iso-configurational ensemble simulations, where a number of trajectories are obtained from the same starting configuration, but with different initial velocities.  The dynamic propensity is defined as either the absolute displacement \cite{bapst2020unveiling, boattini2020autonomously} or the bond-breaking correlation function \cite{jung2022predicting, jung2023roadmap} of a particle in a specified time interval, averaged over all the trajectories in the ensemble.  Simulations clearly show dynamic heterogeneity, i.e., there is a distribution of propensities, with particles of similar propensities clustered spatially.  The simulations do not, however, elucidate the origin of this heterogeneity.

Machine learning (ML) has become a powerful tool in computational physics, and there have been many attempts to apply these techniques to investigate the dynamic heterogeneity of supercooled liquids \cite{boattini2021averaging,jung2022predicting, jung2023roadmap,oyama2023deep,alkemade2022comparing, bapst2020unveiling, cubuk2015identifying, schoenholz2016structural,PhysRevLett.127.088007}.
The majority of studies are supervised methods where the model is trained on a particular output label, e.g., the propensity.
A drawback of supervised methods is that they require prior knowledge of the dynamic propensities of the training dataset, which necessitates long-time iso-configurational ensemble simulations.  In addition, the large number of fit parameters necessitates a substantial amount of training data\cite{bapst2020unveiling}, making supervised methods computationally intensive.  They are also not generalizable, i.e., for every new system, a new training dataset must be generated.   There have been a few unsupervised ML studies (which do not require prior training with target properties), but their performance is less robust than the supervised methods\cite{boattini2020autonomously, paret2020assessing, coslovich2022dimensionality, jung2023roadmap}.

\begin{figure*} 
    \centering
    \includegraphics[width=0.93\textwidth]{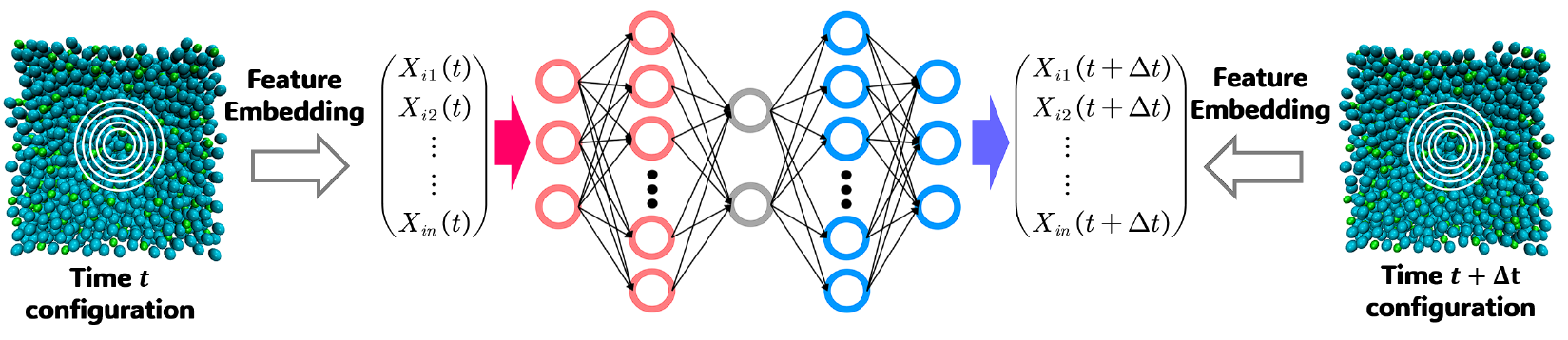}
    \caption{Schematic representation of the unsupervised machine learning method with a Time-lagged AutoEncoder (TAE).  At a given temperature, the input feature vector ${\bf X}_i(t)$ consists of the local structure of the $i^{th}$ particle at time $t$, while the output feature vector ${\bf X}_i(t+\Delta t)$ consists of the same same features, for the same particle, at time $t + \Delta t$. After training, the order parameters are the values of the latent space (grey) variables.}
    \label{TAE_pipeline}
\end{figure*}

In this work we investigate dynamic heterogeneity of supercooled liquids via unsupervised ML.  The departure from previous unsupervised methods is that we use a time-lagged autoencoder (TAE)\cite{wehmeyer2018time} (Figure.~\ref{TAE_pipeline}). The TAE is a neural network that reconstructs the structural features of each particle at time $t+\Delta t$ from its corresponding features at time $t$. The latent space variables of each particle serve as order parameters.  We find that these order parameters correlate with the propensity, and therefore the width of the distribution of order parameters is a measure of the dynamic heterogeneity.  Note that the training does not use any information about the target property, e.g., propensity, and it is therefore completely unsupervised.

A key parameter in the method is the lag time, $\Delta t$.  If $\Delta t$=0, the TAE reduces to the autoencoder, whose order parameters show limited correlation with the long-time dynamics\cite{boattini2020autonomously, coslovich2022dimensionality}.  If $\Delta t$ is large, then the method is not useful because it requires long-time simulations.  We demonstrate that for small but non-zero $\Delta t$, e.g., one thousandth the relaxation time, $\tau$ (the time where self-part of intermediate scattering function decays to $1/e$ of the initial value), the TAE order parameter shows a strong correlation with the propensity at long times.

We study the 3D Kob-Andersen (KA) 80:20 binary Lennard-Jones mixture \cite{kob1995testing}.  The system consists of 3277 particles of type A and 819 particles of type B, which interact via a Lennard–Jones potential:
$    V_{\alpha\beta}(r)=4\epsilon_{\alpha\beta}\left[\left(\frac{\sigma_{\alpha\beta}}{r}\right)^{12}-\left(\frac{\sigma_{\alpha\beta}}{r}\right)^6\right]
$
where $\alpha,\beta\in\{A, B\}$,  $\epsilon_{AA}=1.0$, $\epsilon_{AB}=1.5$, $\epsilon_{BB}=0.5$, $\sigma_{AA}=1.0$, $\sigma_{AB}=0.8$ and $\sigma_{BB}=0.88$.  The units for distance, time, and temperature are $\sigma_{AA}$, $\sigma_{AA}\sqrt{m/\epsilon_{AA}}$, and $\epsilon_{AA}/k_{B}$, respectively, where $k_{B}$ is Boltzmann's constant, and $m$ is the mass of the particles.
Equilibrium configurations are obtained for reduced temperatures ranging from $0.44$ to $0.56$.  Dynamic propensities 
are calculated from at least 30 independent isoconfigurational ensemble simulations (see the Supplemental Material (SM) for details). All results are reported for A-particles, but we note that all findings are independent of particle type. We also characterize the dynamics by the non-Gaussian parameter, $\alpha(t) \equiv \frac{3}{5}\frac{\langle (\Delta r)^{4}\rangle}{\langle (\Delta r)^{2}\rangle^{2}}-1$ where $\Delta r$ is the particle displacement at time $t$. The peak value of $\alpha$ is denoted as $\alpha_m$.

The first step in unsupervised ML methods is the construction of the feature vector ${\bf X}_{i}(t)$ for particle $i$ at time $t$.  Following previous work\cite{boattini2021averaging}, we employ the radial density distribution around particle $i$, which is expressed through Gaussian kernel functions:
$
G_{i}(r, \delta, s)=\sum_{j\neq i, s_{j}=s}e^{-\frac{(r_{ij}-r)^{2}}{2\delta^{2}}}
    \label{radial_features}
$,
where $r_{ij}$ signifies the distance between particle $i$ and its surrounding particle $j$, $s_{j} = \{A, B\}$ is the species of particle $j$. We define $r$ by considering 60 radial shells between 0.5 and 2 (in units of $\sigma_{AA}$, with  $\delta=0.025$), 20 between 2 and 3 (with $\delta=0.05$), and 20 between 3 and 5 (with $\delta=0.1$).  The value of $G_{i}$ in each shell of two particle types constitute a 200-dimensional vector.  The $G_{i}$ vector is processed via mean-free and covariance matrix whitening is processed (see SM) to give the feature ${\bf X}_{i}$ for particle $i$.  We also follow previous studies to use rotationally-invariant spherical harmonics functions to construct angular features but find that they do not provide useful information (see SM).

The TAE network (Figure \ref{TAE_pipeline}) is constructed to map ${\bf X}_{i}(t)$(input) to ${\bf X}_{i}(t+\Delta t)$(output).  Once the latent space variables are identified, we utilize principal component analysis (PCA) to orthogonalize and re-order the them, obtaining two independent order parameters denoted as $\lambda_1$ and $\lambda_2$ (details are in the SM).

\begin{figure}
    \includegraphics[width=0.51\textwidth]{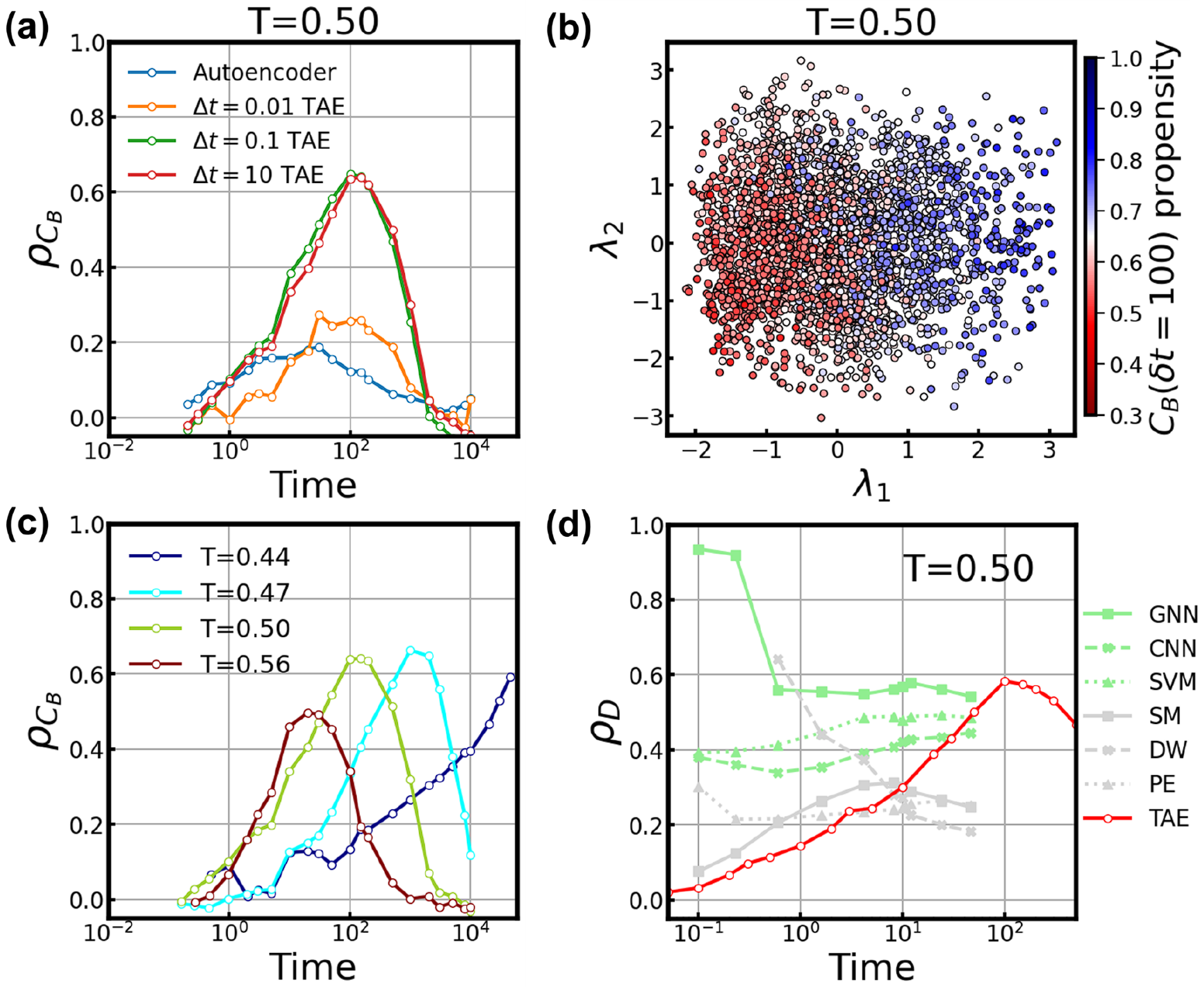}
    \caption{Correlation between dynamic propensity and order parameter. 
    (a)  The Pearson correlation between bond-breaking propensity $C_{B}$ and $\lambda_1$, obtained from TAE with $\Delta t =$ 0.01, 0.1, and 10 for A-particles at temperature $T=0.50$. The correlation between $C_{B}$ and the order parameter from AE with $\Delta t=0$ is shown for comparison.
    (b) The 2-dimensional latent space in the TAE ($\Delta t =0.1$). Each point represents an individual A-particle, and the color is assigned based on the $C_{B}$ propensity. 
    (c) Correlation between $C_{B}$ and the order parameter $\lambda_{1}$ of TAE  models ($\Delta t=0.1$) constructed at different temperatures.
    (d) Comparison with results from supervised machine learning methods and physics-based order parameters taken from Ref. \onlinecite{bapst2020unveiling}. In this case, the correlation is computed based on the absolute displacement propensity.
    The relaxation times for T=0.44, 0.47, 0.50, and 0.56, are $\tau$=4961, 77, 10, and 2, respectively.
    }\label{correlate_propensities}
\end{figure}

We find the latent order parameter $\lambda_1$ correlates well with long-time propensities from isoconfigurational simulations.  Figure \ref{correlate_propensities} (a) illustrates the Pearson correlation coefficient $\rho_{C_{B}}=cov(C_{B}^{i}, \lambda_{1}^{i})/\sqrt{var(C_{B}^{i})var(\lambda_{1}^{i})}$ for T=0.50, where $C_B^i$ is the bond-breaking propensity of particle $i$, and $\lambda^i_{1}$ is the TAE order parameter of particle $i$. Even with a small lag time  $\Delta t=0.1 (\approx 0.01\tau)$, $\lambda_1$ exhibits a strong correlation with bond-breaking propensities from  time $\sim\tau$ to $\sim50\tau$.  In contrast, for $\Delta t=0$ (autoencoder) or 0.01, there is only a weak correlation, and there is no difference in the correlation between $\Delta t$=0.1 and 10.
This suggests that static fluctuations at fairly short times capture long-time dynamic heterogeneity. Figure \ref{correlate_propensities}(b) 
shows a map of the two order parameters where each point represents a particle color coded according to its propensity.  There is a clustering of slow and fast particles, along the $\lambda_1$ coordinate, although the separation is not sharp.  The TAE demonstrates robust performance across a wide temperature range and, the peak in the correlation coefficient occurs around the time-scale of the relaxation time of the system (Figure \ref{correlate_propensities}(c)).  
As the temperature is decreased, the TAE order parameter exhibits strong correlation with propensity over longer time scales. 

The performance of the TAE is comparable to those of supervised ML models. Figure \ref{correlate_propensities}(d) compares the TAE to previous supervised models, including graph neural network (GNN), convolutional neural network (CNN) and support vector machine (SVM), in terms of  the Pearson correlation coefficient $\rho_{D}$ between $\lambda_1$ and the absolute displacement propensity \cite{bapst2020unveiling}. The TAE predictions are comparable or slightly superior to supervised models in the long-time regime. This is significant because the supervised methods are trained on the very quantity, i.e., propensity, that they attempt to predict, whereas the TAE predictions rely solely on the underlying structures.  Additionally, we compare the soft modes (SM) method\cite{widmer2008irreversible}, which utilizes the mode participation fraction of each particle for the low-frequency soft normal modes, the Debye-Waller (DW) factor\cite{widmer2006predicting} that employs the ground-truth dynamics up to $3/4$ of the initial value of the intermediate scattering function (i.e.,corresponds to time $\approx 0.4$ for $T=0.50$), and the potential energy (PE)\cite{berthier2007structure, doliwa2003does} as order parameters to discern long-time dynamic heterogeneity. The TAE has a higher Pearson correlation coefficient for long-time dynamics compared to these approaches.

\begin{figure}
    \includegraphics[width=0.47\textwidth]{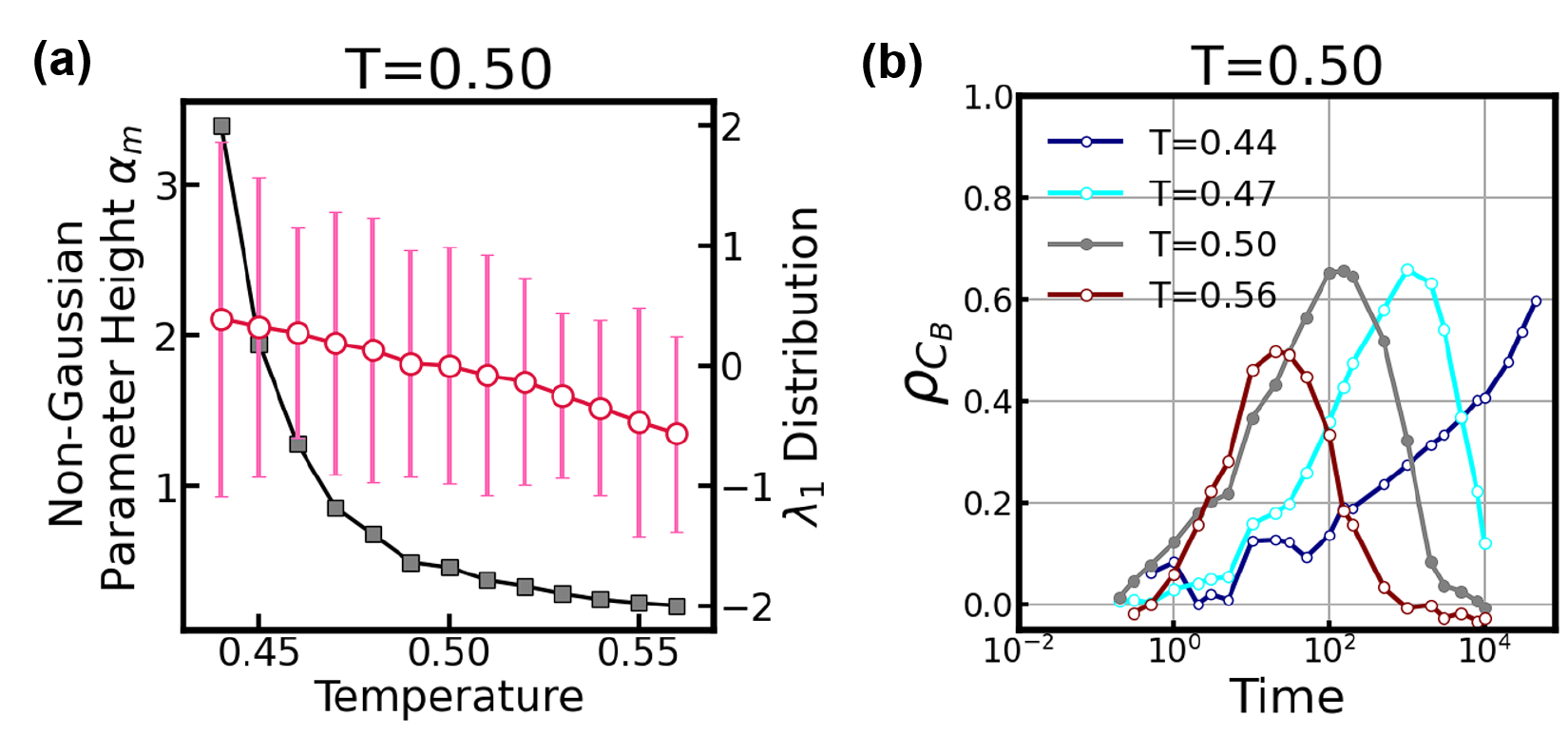}
    \caption{(a) The distribution of $\lambda_1$ (red dots) at different temperatures encoded by a TAE constructed at $T=0.50$.
    The heights, $\alpha_m$, of the peaks of the non-Gaussian parameter (black squares) are shown for comparison. (b) Transferability of the TAE to different temperatures.  The TAE is constructed at T=0.5 and used in conjunction with configurations at other temperatures to obtain $\lambda_{1}$ at that temperature.  The correlation with propensities obtained at the corresponding temperature are shown.}
    \label{non_gaussian_relation}
\end{figure}

After optimizing the parameters (i.e., weights and biases) of the TAE at a specific temperature, they can be transferred to other temperatures without additional optimization. To evaluate the transferability of the TAE,  we construct the network at one temperature, e.g., T=0.50, and then use this to encode features of particles from equilibrium configurations at other temperatures, thus determining the order parameter value as a function of temperature (Note that this does not require additional simulations at other temperatures, just ensembles of configurations from equilibrium simulations). At each of the other temperatures, there is a distribution of $\lambda_1$ due to A-particles from different configurations.  Figure \ref{non_gaussian_relation}(a) shows that the TAE predicts an increase in $\lambda_1$ as the temperature is decreased.  The increase in $\lambda_1$ is linear and not as dramatic and sensitive as in the case of, for example, the peak value of non-Gaussian parameter $\alpha_{m}$, shown for comparison. The values of $\lambda_1$ thus obtained, however, show good correlation with the propensities measured at the corresponding temperatures (Figure \ref{non_gaussian_relation} (b)). The predictive performance of the TAE at different temperatures is comparable to that of TAE specifically constructed for those temperatures (Figure \ref{correlate_propensities}(c)). The TAE therefore has predictive power at temperatures different from where its parameters are determined.

\begin{figure}
    \includegraphics[width=0.42\textwidth]{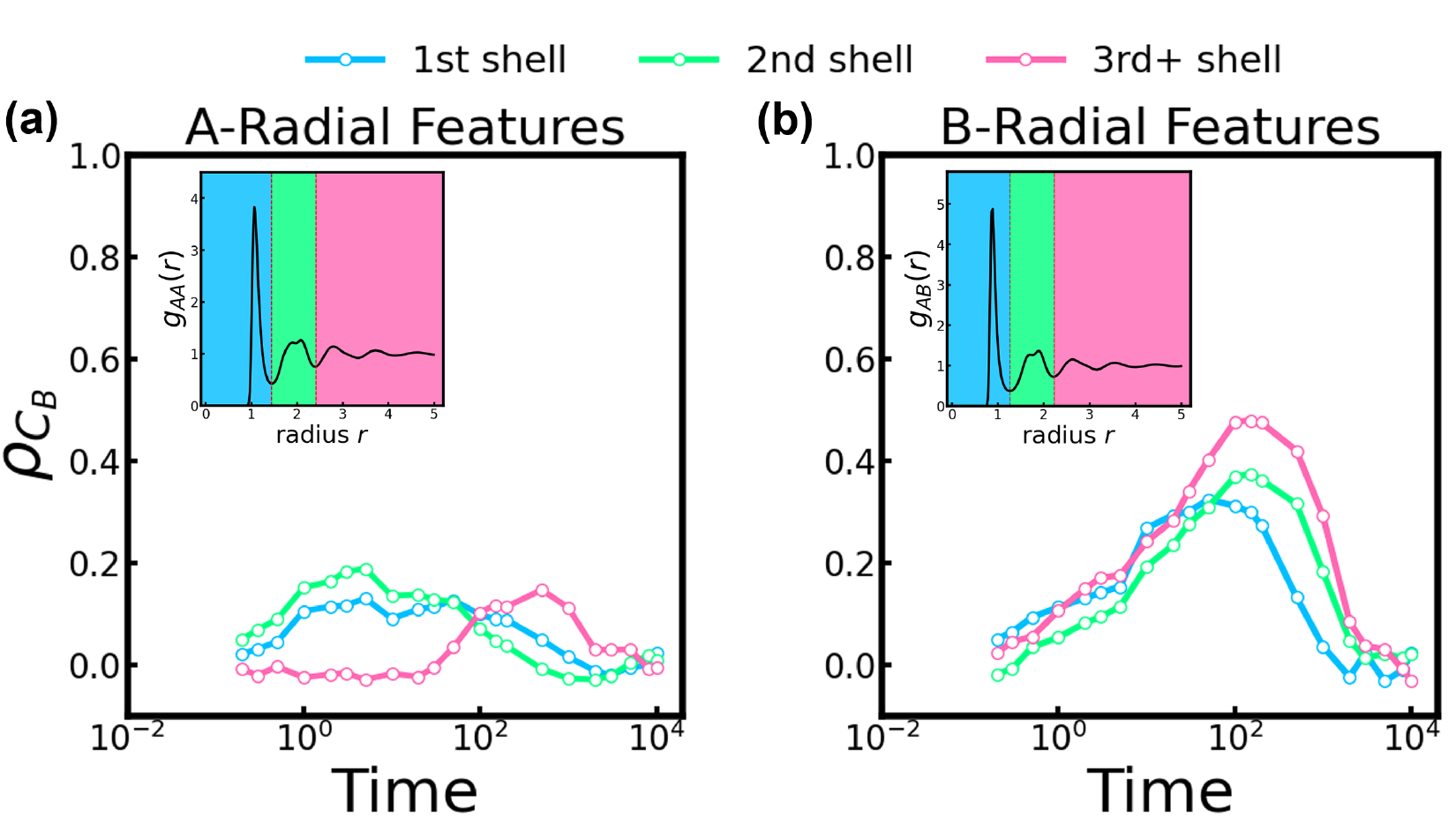}
    \caption{The radial density features subsampled based on the particle types (a) A-type and (b) B-type, and for different spatial  regions. The different curves for the correlation between $\lambda_1$ and the propensity are obtained using features from similarly colored regions in the graph for the pair correlation function.}
    \label{feature_importance}
\end{figure}

Radial features at all length-scales are important for the success of the TAE. 
To investigate the importance of features for dynamic prediction, we restrict the TAE (T=0.50, $\Delta t=0.1$) to features in three regions: from 0 to the first minimum in the pair correlation function, between the first and second minima, and beyond the second minimum. (Figure \ref{feature_importance}). When only the densities of surrounding A-particles are used in the feature (Figure \ref{feature_importance}(a)), the correlation with propensity is limited; however, when only the densities of surrounding B-particles are used (Figure \ref{feature_importance}(b)), the correlation with propensity is stronger.  Radial features on all length-scales are required for a good correlation with the propensity;  the performance of the TAE with a subset of features (Figure \ref{feature_importance}) is not as strong as when all the features are employed (Figure \ref{correlate_propensities} (a)).  This is consistent with a recent unsupervised ML study on the phases of water 
\cite{doi:10.1021/acs.jpclett.4c00383} and a supervised ML study on the KA model \cite{PhysRevLett.127.088007}.

In summary, we introduce a fast and easily implementable unsupervised ML method to study the dynamic heterogeneity of supercooled liquids.  The model identifies an order parameter, based on structural information, that correlates well with dynamic propensity. Using information from short simulations at a specific temperature to construct TAE, the order parameter can be transferred to correlate with long-term propensities at other temperatures. Importantly the TAE uses a single time lag, which is short enough to be computationally feasible and requires no information about the target property.  An intriguing conclusion is that the essence of the long time dynamics is already encoded into the fluctuating structural behaviors at short times, although the ``structure" is not a simple pair-based metric but rather arises from a non-linear transformation through the neural network.    The results also demonstrate the utility of unsupervised ML models in the study of dynamic processes in liquids, which is currently an unexplored area of physics.

\acknowledgments  

We thanks M. Ediger and Y. Wang for fruitful discussions. X.H. acknowledges the support from the Hirschfelder Professorship Fund. We acknowledge computational resource support from the Center for High Throughput Computing at the University of Wisconsin-Madison. The codes for time-lagged analysis of supercooled liquid dynamics can be accessed for public use on https://github.com/YunruiQIU/supercooled-dynamics.

\bibliographystyle{unsrt}
\bibliography{reference}

\newpage

\newpage



\section*{Supplementary material}

# Supplementary Material: Investigate dynamic heterogeneity of supercooled liquid by unsupervised machine learning

Yunrui Qiu,[1] Inhyuk Jang,[1] Xuhui Huang*,[1,2] and Arun Yethiraj[†1]

[1]*Department of Chemistry, Theoretical Chemistry Institute,*
*University of Wisconsin-Madison, Madison, Wisconsin 53706, United States*
[2]*Data Science Institute, University of Wisconsin-Madison, Madison, Wisconsin 53706, United States*
(Dated: April 23, 2024)

**Contents**

---

\* Corresponding author:xhuang@chem.wisc.edu
† Corresponding author:yethiraj@wisc.edu

# I. MODEL AND SIMULATION METHOD

## A. Kob-Andersen Model

We perform the molecular dynamics (MD) simulations under various temperatures and pressures for a system comprising an 80:20 binary mixture, with a total of $N = 4096$ particles within a three-dimensional cubic box. All the simulations are conducted using LAMMPS[1] package. The 3277 large particles of A type and 819 small particles of B type interact with each others via 6-12 Lennard–Jones potential:

$$V_{\alpha\beta}(r) = 4\epsilon_{\alpha\beta}[(\frac{\sigma_{\alpha\beta}}{r})^{12} - (\frac{\sigma_{\alpha\beta}}{r})^6] \tag{1}$$

where $\alpha, \beta \in \{A, B\}$. We adopt the parameters in potential based on the Kob-Andersen configuration[2] where $\epsilon_{AA} = 1.0$, $\epsilon_{AB} = 1.5$, $\epsilon_{BB} = 0.5$, $\sigma_{AA} = 1.0$, $\sigma_{AB} = 0.8$ and $\sigma_{BB} = 0.88$. The long-range interaction is truncated for distances greater than $r_c = 2.5\sigma_{AA}$, and periodic boundary conditions are employed. Both types of particles are assumed to have the identical mass $m$. Based on the Lennard-Jones potential, we establish dimensionless units for distances, time, temperature, and pressure as $\sigma_{AA}$, $\sqrt{\frac{m\sigma^2 AA}{\epsilon AA}}$, $\epsilon_{AA}/k_B$ (where $k_B$ represents the Boltzmann constant), and $\epsilon_{AA}/\sigma^3_{AA}$, respectively.

We conduct MD simulations covering the temperature range from 0.44 to 0.56 with the increment of 0.1, and the pressure range from 0.17 to 2.93 with the increment of 0.23. The simulations for each state point $(T, P, \rho)$ can be divided into five steps[3]: (i) Randomly initializing coordinates for all particles and performing energy minimization. (ii) Equilibrating the system at an initial high-temperature state with $T_h = 0.5495$, $P_s = 0.4888$ using the *NPT* Nose-Hoover thermostat. (iii) Performing constant *NPT* equilibration run at target temperature range from 0.44 to 0.56 and target pressure range from 0.17 to 2.93 using Nose Hoover thermostat. The equilibration duration at each temperature is ten times longer than the corresponding relaxation timescale. (iv) Carrying out a continuous *NVT* equilibration run at the target temperature, lasting ten times longer than the corresponding relaxation timescale. (v) Executing a continuous *NVE* ensemble production run, employing the Velocity-Verlet integrator.

To characterize the dynamic properties of the 3D Kob-Andersen system, we compute ensemble averages of mean squared displacement, non-Gaussian parameter, self-intermediate scattering function, and relaxation time for simulations at different temperatures and pressures (Figure 1).

## B. Isoconfigurational Simulation

To evaluate the dynamic heterogeneity of a supercooled liquid, propensity serves as a representative feature that measures the mobility of each particle over different time intervals, dependent solely on the initial configuration. We perform isoconfigurational simulations for each specific initial configuration and calculate the ensemble average of the bond-breaking correlation function[4, 5] or absolute displacement[6, 7] of each particle to determine its dynamic propensity. For each configuration, we conduct at least 30 independent simulations, initializing particle velocities randomly from the Maxwell-Boltzmann distribution at the corresponding temperature. The ensemble average bond-breaking correlation propensity of particle $i$ after a time interval $\delta t$ is defined as:

$$C_B^i(\delta t) = \langle n_{\delta t}^i / n_0^i \rangle_{iso} \tag{2}$$

where $n_0^i$ represents the number of neighboring particles $j$ within a cutoff of $r_{cut}^0 = 1.4\sigma_{\alpha_i\beta_j}$ for particle $i$ at time 0, and $n_{\delta t}^i$ denotes the number of particles that were initial neighbors and still remain within a cutoff of $r_{cut}^{\delta t} = 1.8\sigma_{\alpha_i\beta_j}$ for particle $i$ after time $\delta t$. $\alpha_i$ and $\beta_j$ represent the type of particle $i$ and $j$, respectively. The initial cutoff $r_{cut}^0$ is determined based on the position of the first peak in the pair correlation function, while the second cutoff $r_{cut}^{\delta t}$ is set to a larger value to ensure that particles truly escape their confinement rather than experiencing minor fluctuations. By definition, the value of $C_B^i(\delta t)$ lies within the range of 0 and 1.

The absolute displacement based propensity of particle $i$ after a time interval $\delta t$ is defined as follows:

$$D_i(\delta t) = \langle |\boldsymbol{r}_i(\delta t) - \boldsymbol{r}_i(0)| \rangle_{iso} \tag{3}$$

where $\boldsymbol{r}_i(t)$ is the position of particle $i$ at time $t$ and the average is taken for all independent isoconfigurational trajectories. In our implementation, we observe that these two different propensities exhibit similar and consistent behaviors when correlated with the order parameters derived from TAE, with the bond-breaking propensity demonstrating a slightly higher correlation (shown in Figure 2 and Figure 3). The dynamical heterogeneity (i.e., the variance of the propensity distribution) increases when temperature decreases (Figure 3).

## II. LOCAL STRUCTURE DESCRIPTORS

### A. Radial Feature Embedding

To characterize the local radial environment of each particle, we utilize the continuous Gaussian kernel function to incorporate the radial features, which has been developed and demonstrated to be highly effective in prior studies[6–8]. The radial features are defined as:

$$G_i(r, \delta, s) = \sum_{j \neq i, s_j = s} e^{-\frac{(r_{ij} - r)^2}{2\delta^2}} \tag{4}$$

where $i$ represents the probe particle for which we aim to assess the radial density, $r_{ij}$ signifies the distance between particle $i$ and its neighbor $j$, $s_j = \{A, B\}$ is the species of particle $j$. By adjusting the values of $r$, $\delta$, and $s$, the Gaussian kernel functions capture different facets of the local density surrounding particle $i$.

We define the values for $r$ and $\delta$ in order to capture the local density within three distinct shells surrounding the probe particle. For the first shell, we evenly divide the range $(0.5\sigma_{AA}, 2.0\sigma_{AA}]$ into 60 intervals to serve as the mean $r$ for sub-shells, and we set $\delta$ to 0.025 for all sub-shells. For the second shell, we uniformly divide the range $(2.0\sigma_{AA}, 3.0\sigma_{AA}]$ into 20 intervals for the mean $r$ of sub-shells, with $\delta$ set to 0.05 for all sub-shells. In the case of the third shell, we uniformly divide the range $(3.0\sigma_{AA}, 5.0\sigma_{AA}]$ into 20 intervals to determine the mean $r$ of sub-shells, and $\delta$ is set to 0.1 for all sub-shells. Consequently, for each probe particle, we encode its local radial distribution of a specific particle type into a 100-dimensional feature vector. Overall, we concatenate radial features from two distinct particle types to create a 200-dimensional radial embedding for each probe particle. All the hyperparameters $r$ and $\delta$ are selected based on previous studies [7, 9, 10].

### B. Angle Feature Embedding

To encode the local angular structural information for particles, we draw inspiration from previous studies[7, 9, 11] and utilize the distance-dependent expansion of the local density in terms of spherical harmonics functions. First, for any given probe particle $i$, we define the complex quantities:

$$q_i(l, m, r, \delta) = \frac{1}{Z} \sum_{j \neq i} e^{-\frac{(r_{ij} - r)^2}{2\delta^2}} Y_l^m(\boldsymbol{r_{ij}}) \tag{5}$$

where $Y_l^m$ represents the spherical harmonic function of order $l$, where $m$ is an integer ranging from $-l$ to $+l$. $\boldsymbol{r_{ij}}$ is the distance vector between particle $i$ and its neighbors $j$ which can be used further to determine polar and azimuthal angle. $Z = \sum_{j \neq i} e^{-\frac{(r_{ij} - r)^2}{2\delta^2}}$ is a normalization constant. Furthermore, rotationally-invariant angular descriptors are defined by doing summation over $m$ as:

$$q_i(l, r, \delta) = \sqrt{\frac{4\pi}{2l+1} \sum_{m=-l}^{m=l} |q_i(l, m, r, \delta)|^2} \tag{6}$$

These measures quantify the $l$-fold symmetry within the distribution of neighboring particles located at a distance $r$ from a probe particle $i$ within a shell of width $2\delta$. To account for various symmetries, we allow $l$ to range all integers between 1 and 12, divide the interval $[1, 2.5]$ into 16 intervals for the radial mean $r$, while maintaining $\delta$ at 0.1. Consequently, we obtain a 192-dimensional feature for each probe particle, representing its local angular information. We truncate the evaluation of angle features for distances $r \leq 2.5\sigma_{AA}$ since longer range of angular distributions become more homogeneous and computationally expensive to embed.

## III. TIME-LAGGED AUTOENCODER (TAE)

The time-lagged autoencoder (TAE) is employed to approximate the dynamical propagator, allowing it to map the current observations to the future observations using low-dimensional representations. It was initially introduced and applied for investigating the dynamics of protein conformational changes.[12] Here, we extend its application to

analyze the dynamics of supercooled liquid. Similar methods exist in other fields, such as dynamic mode decomposition (DMD) [13, 14] and canonical correlation analysis (CCA).[15, 16]

The TAE can identify the optimal set of non-linear-transformed and low-dimensional variables, where the projection of the time-lagged data from the dynamical system exhibits the highest time-lagged correlation. These low-dimensional variables are considered the most important and predictive order parameters for forecasting the system's dynamics and they correspond to the slowest fluctuating degrees of freedom in the data space. In our implementation, we utilize time-lagged features which describe the local structural information of particles in two time-lagged configurations as the input and output to optimize the TAE, respectively. The schematic architecture of the implemented TAE is presented in the main text and Figure 4.

The training data for TAE can be represented as follows: $\{\boldsymbol{X}_i(t), \boldsymbol{X}_i(t + \Delta t)\}_{i=1}^N$, where $\boldsymbol{X}_i(t) \in \mathbb{R}^d$ is the feature for particle $i$ embedded at time $t$ with $d$ dimensions, and $\boldsymbol{X}_i(t + \Delta t) \in \mathbb{R}^d$ corresponds to the feature for the same particle $i$ at time $t + \Delta t$. The dimension is $d = 200$ for the radial-features-based TAE, while it is $d = 192$ for the TAE based on angular features. $N$ represents the number of particles for which time-lagged features are utilized in training. For the TAE model presented in this paper at a specific temperature, we used 100 independent pairs of equilibrium configurations (each contains 3722 A-particles) at the specified temperature for the training of the model. But we observe that the number of training time-lagged configurations could be much less to achieve the same performance. To evaluate the correlation between TAE order parameters and dynamic propensity, 30 configurations are randomly selected from the 100 equilibrium configurations, and for each selected configuration, at least 30 independent isoconfigurational simulations are conducted to obtain propensity at different times for every particle.

We construct the AutoEncoder network for TAE with a single nonlinear-layer encoder and a single nonlinear-layer decoder, both utilizing the Tanh activation function. The number of neurons in the hidden layers is consistently set as five times the input dimensions, i.e., $5d$. For the bottleneck and output layers, we utilize a linear activation function. The Adam optimization method is employed to optimize the TAE parameters, and the StepLR scheduler is used to decay the learning rate. The detailed data processing procedure is:

- (1). Move the mean and whiten the feature data:

$$\tilde{\boldsymbol{X}}_i(t) = \mathbf{C}_{00}^{-\frac{1}{2}}(\boldsymbol{X}_i(t) - \frac{1}{N}\sum_{i=1}^N \boldsymbol{X}_i(t)) \tag{7}$$

$$\tilde{\boldsymbol{X}_i}(t + \Delta t) = \mathbf{C}_{\Delta t \Delta t}^{-\frac{1}{2}}(\boldsymbol{X}_i(t + \Delta t) - \frac{1}{N}\sum_{i=1}^N \boldsymbol{X}_i(t + \Delta t)) \tag{8}$$

where $\mathbf{C}_{00}$ and $\mathbf{C}_{\Delta t \Delta t}$ are self-covariance matrices for features:

$$\mathbf{C}_{00} = \frac{1}{N}\sum_{i=1}^N \boldsymbol{X}_i(t)\boldsymbol{X}_i^T(t) \tag{9}$$

$$\mathbf{C}_{\Delta t \Delta t} = \frac{1}{N}\sum_{i=1}^N \boldsymbol{X}_i(t + \Delta t)\boldsymbol{X}_i^T(t + \Delta t) \tag{10}$$

- (2). Update the TAE parameters based on loss function defined as:

$$\mathcal{L} = \frac{1}{N}\sum_{i=1}^N ||\tilde{\boldsymbol{X}_i}(t + \Delta t)) - \mathbb{D}(\mathbb{E}\tilde{\boldsymbol{X}_i}(t))||_2 + \lambda \sum_{k=1}^M \omega_k^2 \tag{11}$$

where $\mathbb{E}$ and $\mathbb{D}$ represent the encoder and decoder neural network, respectively. And $\{\omega_k\}_{k=1}^M$ represents the set that includes all weight parameters in the neural network.

- (3). Employ Principal Component Analysis (PCA) to re-order and orthogonalize the TAE latent space, setting the first component as the order parameter $\lambda_1$. Then derive order parameters for particles by encoding their features using the optimized encoder and PCA. It is crucial to establish whether the latent variables and dynamics propensities exhibit positive or negative correlation in advance. This can be achieved by encoding particles with known propensities or encoding particles from different temperatures. Once the relations are determined, order parameters can be fixed and generalized.

It is noted that while the number of hidden layers and the number of neurons in hidden layers are adjustable, the TAE consistently shows robust performance across various setups. And we consistently utilize a two-dimensional bottleneck space in the TAE architecture, as we observe that increasing dimensions of latent space does not enhance the performance of dynamics prediction (shown later).

Additionally, it's worth mentioning that replacing the non-linear AutoEncoder with a linear matrix allows us to use linear regression for identifying the optimal linear propagator. This aligns with the time-lagged canonical correlation analysis (TCCA) method, where the top singular functions can be regarded as the order parameters. We observe that TCCA also demonstrates the ability to identify order parameters highly correlated with long-term propensity.

## IV. SUPPLEMENTARY NOTE 1: FEATURE IMPORTANCE

### A. Comparison of Radial and Angular Features

To employ unsupervised machine learning methods for discerning dynamic heterogeneity of supercooled liquid, the initial step always involves selecting the appropriate structural features to describe the local environment. We compare two types of features which are commonly employed as local descriptors: radial density and angular distribution. We apply three different feature combinations (i.e., only radial, only angular, and combined radial and angular features) to construct two different models, AE and TAE. The constructions utilize equilibrium configurations of A-particles lagged by $\Delta t = 0.1$ at temperatures $T = 0.44$ and $T = 0.50$, respectively. All training procedures are identical to the above descriptions. The performance of order parameters constructed by different features and different models are shown in Figure 5 for $T = 0.44$ system and Figure 6 for $T = 0.50$ system. Indeed, for both systems and two different models, the order parameters generated from the angular features exhibit limited correlation with propensity across a long time range. In contrast, the performance of order parameters constructed from the radial features is comparable to those constructed from the combined features, and both of them are quite effective. Especially, by visualizing dynamical heterogeneity across snapshots of the Kob-Andersen system at different temperatures, we found that the order parameter derived from radial-feature-based TAE clearly distinguishes long-term propensities (Figure 3). Thus, our observations support the notion that radial density primarily correlates with the dynamical behaviors of the 3D Kob-Andersen system, which is consistent with prior research[7, 8, 17]. Additionally, we observe that the displacement occurring during the lag time $\Delta t = 0.1$ exhibits a strong correlation with propensity at $\delta t = 0.1$, but demonstrates no correlation with long-term dynamics. We also note that in the case of other supercooled liquid systems, investigating the dynamics may require a greater emphasis on angular features.

### B. Comparison of Radial Features of Different Shells

In the 3D Kob-Andersen system, we observe that radial features play a more crucial role in predicting dynamics compared to angular features. In this section, we delve deeper into understanding the importance of radial features within shells with different length ranges. By calculating the pair correlation function for A-particle with respect to its surrounding A-particles (denoted as $g_{AA}(r)$) and B-particles (denoted as $g_{AB}(r)$), we can infer the ensemble average locations of different shells which are defined as regions between different locations of the minimum values in the pair correlation function. We subsequently construct multiple independent TAE models using radial features from different shells and examine effectiveness of their order parameter in capturing dynamic propensity.

Specifically, we conduct the analysis by categorizing the 200-dimensional radial feature of A-particle into six regions: three distinct shells (i.e., from 0 to the first minimum in the pair correlation function, between the first and second minima, and beyond the second minimum) concerning two types of particles, as depicted in Figure 7 with different colors. We utilize features from these six regions to construct six TAE models at four different temperatures: $T = 0.44, 0.47, 0.50, 0.56$, respectively. Each model is optimized using 100 pairs of equilibrium configurations lagged by $\Delta t = 0.1$ at the corresponding temperature, and the number of hidden neurons is consistently set to five times the dimensions of the input features. Other setups are identical to those described above. Then the order parameter $\lambda_1$ from the latent spaces of TAEs, trained with features from different shells, are used to discern dynamic heterogeneity.

As illustrated in Figure 7, we observe three notable conclusions:

(1) The radial features concerning the distribution of neighbor B-particles play a more crucial role in understanding the long-time dynamic heterogeneities of probe A-particle. Clearly, although radial features of neighbor A-particles exhibit larger fluctuations, they do not contribute significantly to the understanding of long-time dynamics. This explains why the autoencoder with a lag time $\Delta t = 0$ showed poor performance. Instead, the radial density features

for neighbor B-particles, characterized by less fluctuations, are identified by TAEs as the slowest fluctuating and most predictive features.

(2) For both A-particle and B-particle radial density features, the short-range density (i.e., 1st shell or 2nd shell) is more influential in short-time dynamics, while the medium-range density (i.e., 3rd+ shell) contains more information about long-time dynamics.

(3) Incorporating features from all length-scale regions is crucial for TAE to generate high-quality order parameter. Order parameter derived solely from features within a specific shell may not demonstrate comparable performance to those constructed from features across all shells.

## V. SUPPLEMENTARY NOTE 2: EFFECTS OF LAG TIME AND LATENT SPACE DIMENSION

It is evident that two crucial hyperparameters may affect the TAE model: the lag time $\Delta t$ and the dimensions of latent space. We construct multiple TAE models with varying lag times and evaluate the effectiveness of different order parameters. As illustrated in Figure 8, we build up TAEs with four-dimensional latent space using radial features with different lag times and from four distinct temperatures. Other training setups are same as described above. We observe once lag time is chosen $0.1 < \Delta t < \tau$ ($\tau$ is relaxation time), minimal variations in the performance of the first order parameter $\lambda_1$ across different lag times and different temperatures. However, if $\Delta t$ is set too small, the order parameter $\lambda_1$ cannot be correctly identified; conversely, if $\Delta t$ is set longer than the relaxation time, rearrangement of the initial structure occurs, resulting in a loss of prediction power. These observations suggest that there is no distinct time-separation in the dynamics of supercooled liquid and the effective order parameter $\lambda_1$ can be identified by TAE model constructed with a broad lag time window.

Additionally, for different order parameters, we find that the top two order parameters appear to be the most significant. The first order parameter $\lambda_1$ consistently exhibits the highest correlation with long-term propensity, while the second one displays slightly higher correlation with short-term dynamics. Therefore, we conclude that two order parameters are sufficient, and additional order parameters cannot provide more information. As a result, we consistently utilize the lag time of $\Delta t = 0.1$ and the two-dimensional latent space to construct TAE models.

## VI. SUPPLEMENTARY NOTE 3: MARKOVIAN EMBEDDING OF NON-MARKOVIAN DYNAMICS

We employ TAE to construct a non-linear propagator that maps the features of a particle at time $t$ to the features of the same particle at time $t + \Delta t$. In this approach, we ignore the historical influence before time $t$ and make the strong assumption that the evolution of dynamics in feature space is Markovian. However, the dynamics of a single particle are expected to have a strong memory effect due to its active interactions with multiple surrounding particles. If we solely utilize the coordinates and velocities of the probe particle to construct the feature space, the dynamics in the feature space should be non-Markovian. This explains why the displacement during a short lag time shows no correlation with long-time propensity at all(Figure 5 and 6). The strong memory effect prevents the predictability of long-term dynamics from short-time displacement alone.

In our implementation of TAE, we address this issue by employing the Markovian embedding of non-Markovian dynamics approach.[18, 19] This involves incorporating a substantial amount of additional degrees of freedom, achieved through the consideration of local surrounding structures. The resulting high-dimensional feature space takes into account interactions between the particle and its surroundings, allowing for a better approximation of dynamics as a Markovian process. But since we consider each particle individually and cannot update the evolution of features due to the motions of surrounding particles, the accurate propagation time of dynamics is limited.

To substantiate the above justifications, we evaluate the quality of order parameters obtained from TAE models constructed with varying numbers of features. We fix the TAE lag time at $\Delta t = 0.1$ and randomly sub-sample the 200-dimensional radial features to train TAE models. The number of hidden neurons is consistently set as five times the number of input features. We repeat this process twenty times for a specific number of features. As shown in Figure 9, we observe a substantial decrease in the prediction accuracy of TAE order parameters when the input features are truncated. In contrast, the prediction accuracy of order parameters from the $\Delta t = 0$ Autoencoder does not change significantly. Despite the identification of the radial features for B-particle as important, utilizing only a single shell of these features lead to order parameters with significantly reduced predictive power (Figure 7). These results demonstrate that expanding the feature space enhances the predictability of long-time dynamics for a single particle using a short lag time, aligning with our concept of Markovian embedding of non-Markovian dynamics.

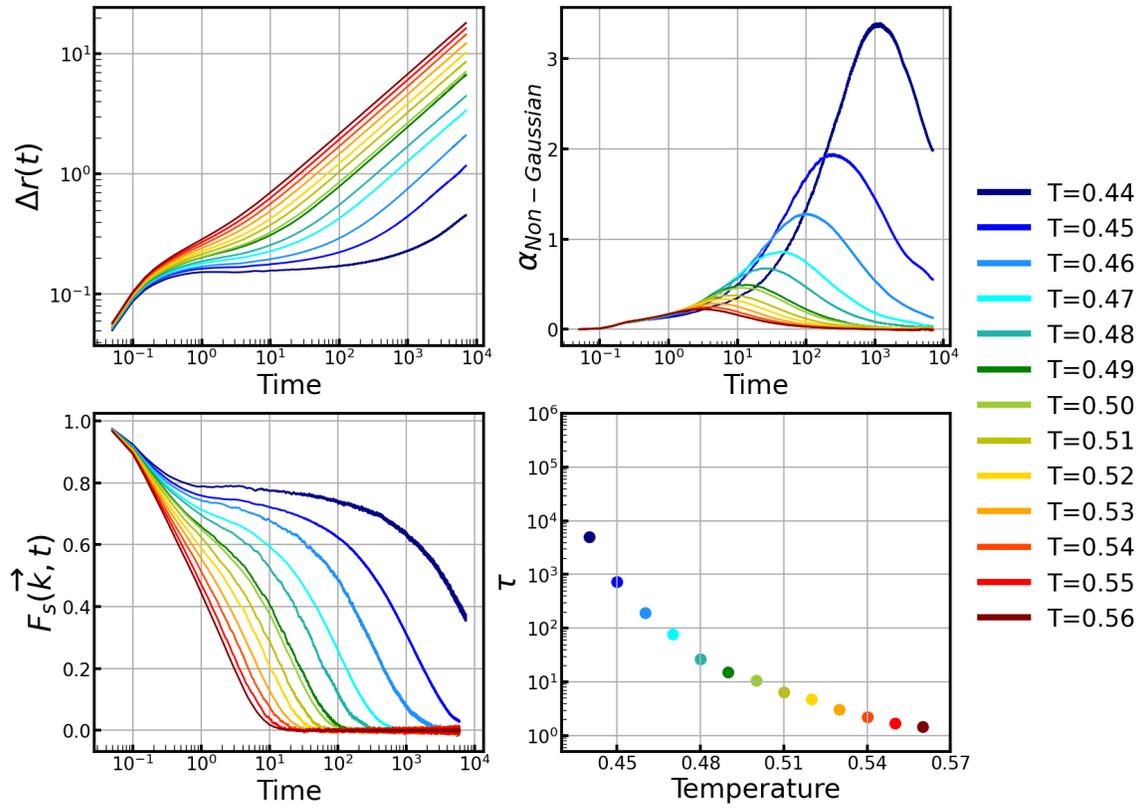

**FIG. 1: Dynamical characteristics of the 3D Kob-Andersen model at different temperatures.** (a) Mean squared displacement (MSD), (b) Non-Gaussian parameter, and (c) Real part of the intermediate scattering function as functions of time at various simulation temperatures. (d) Relaxation timescale $\tau = F_s^{-1}(\frac{1}{e})$ for different temperatures.

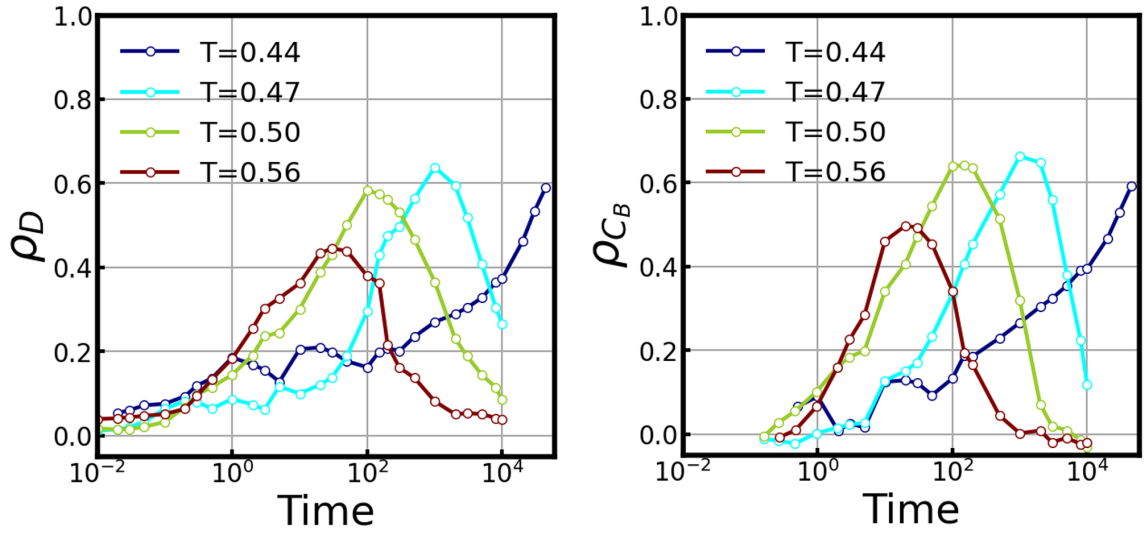

**FIG. 2: Compare the correlations between TAE order parameter and propensities with different definitions.** TAE models are constructed with 200-dimensional radial features from four different temperatures respectively ($\Delta t = 0.1$), and the order parameter $\lambda_1$ is correlated with absolute displacement based propensity $D_i(\delta t)$ and bond-breaking correlation based propensity $C_B^i(\delta t)$.

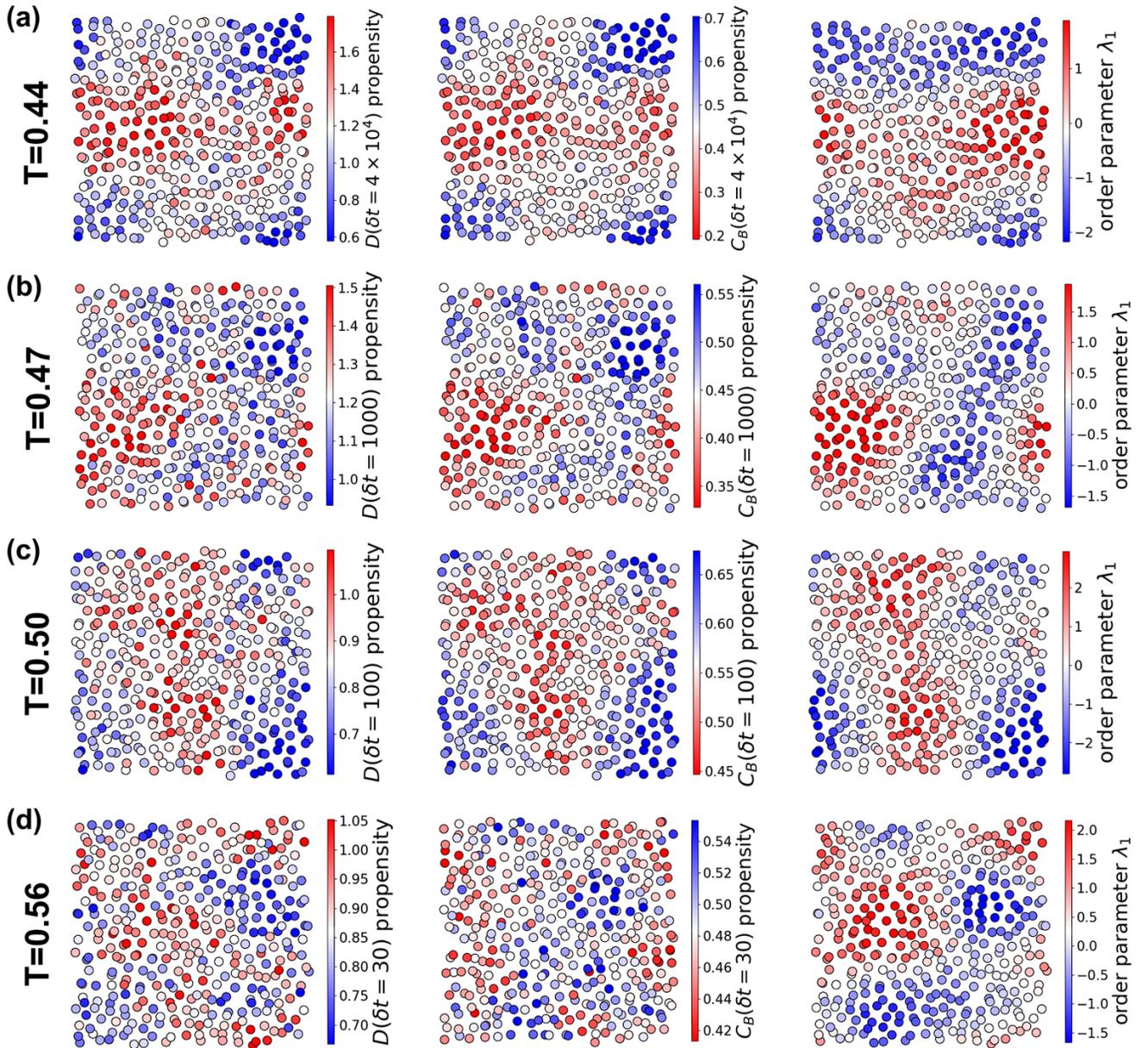

**FIG. 3: Dynamical propensity calculated from isoconfigurational simulations and predicted from order parameter $\lambda_1$ at different temperatures.** Snapshots of representative configuration randomly selected from different temperatures (different rows) at different timescales are presented. Each particle is visualized as a scatter and colored by absolute-displacement propensity (first column), bond-breaking correlation propensity (second column) and value of $\lambda_1$ from TAE (third column). The TAE is trained only with radial features and $\Delta t$ is set as 0.1. The color bar is scaled to cover four times the corresponding variance.

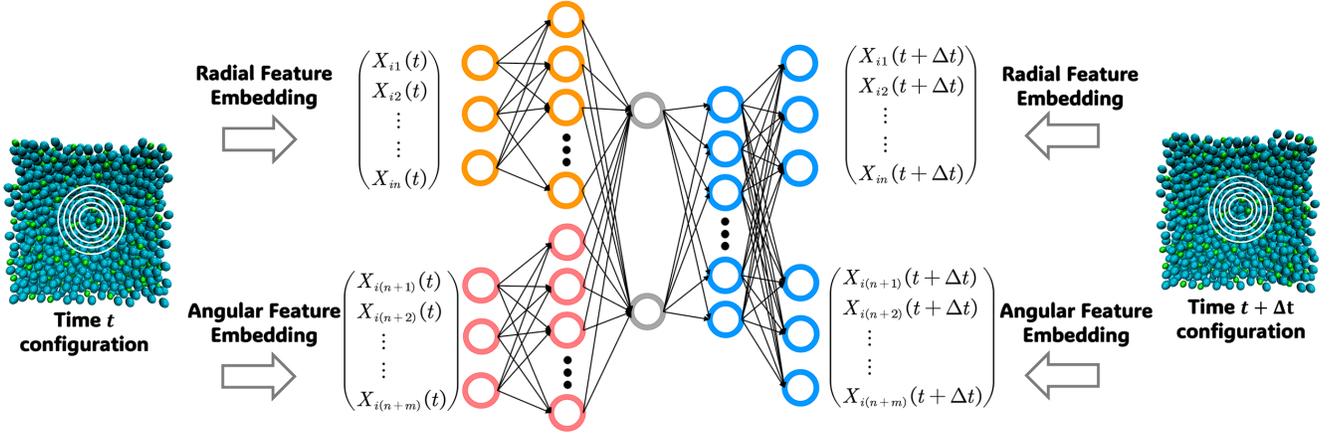

**FIG. 4: Schematic representation of the unsupervised machine learning with Time lagged AutoEncoder (TAE) using radial and angular combined features**. For a given pair of time-lagged equilibrium configurations at a certain temperature, the input feature vector $\mathbf{X}_i(t)$ signifies the features of the $i^{th}$ particle at time $t$, while the output is the feature vector $\mathbf{X}_i(t + \Delta t)$ for the same particle at time $t + \Delta t$. Here, the feature for each particle is constructed by a 392-dimensional radial density and angular distribution combined vector. The encoder (red) and decoder (blue) neural networks are trained using feature of every particle with the same species from the given pairs of time-lagged configurations, and the latent space order parameters are represented by the grey circles.

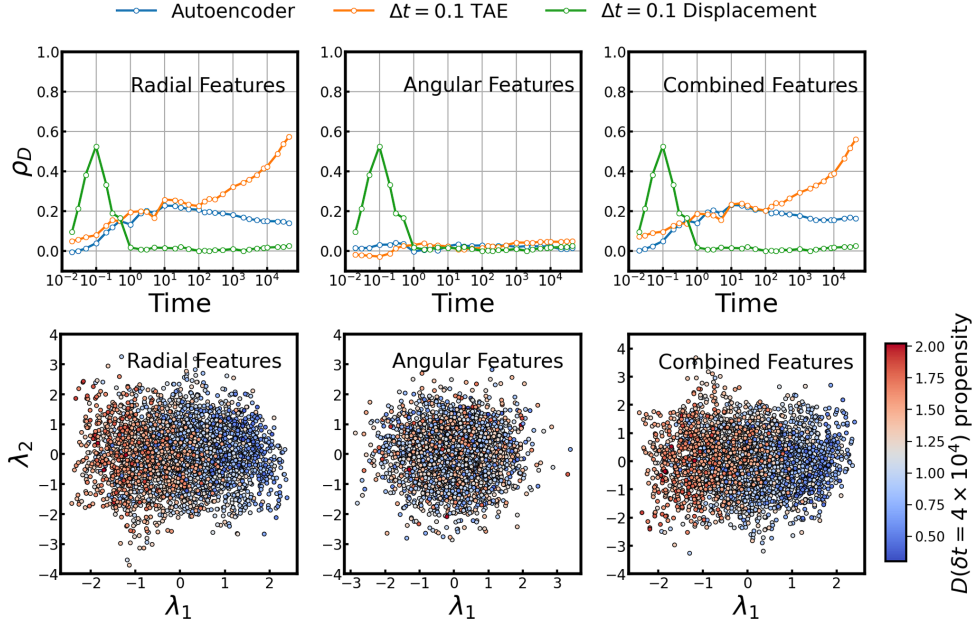

**FIG. 5: Importance of radial and angular features.** The local structures of A-particles from equilibrium configurations at temperature $T = 0.44$ are embedded with different descriptors: radial features, angular features, and combined radial and angular features. Different features are utilized to train AE and TAE ($\Delta t = 0.1$) models, respectively. For each model, the correlation between the order parameter $\lambda_1$ and propensity is visualized. The distributions of particles on two order parameters derived from different TAE models are also visualized, with particles color-coded according to their long-term propensities.

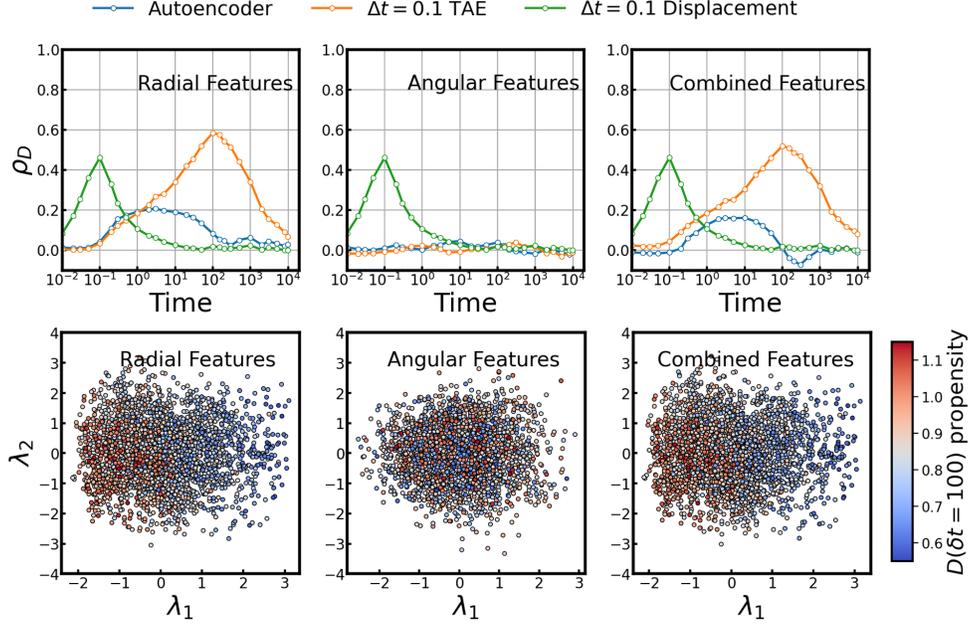

**FIG. 6: Importance of radial and angular features.** The local structures of A-particles from equilibrium configurations at temperature $T = 0.50$ are embedded with different descriptors: radial features, angular features, and combined radial and angular features. Different features are utilized to train AE and TAE ($\Delta t = 0.1$) models, respectively. For each model, the correlation between the order parameter $\lambda_1$ and propensity is visualized. The distributions of particles on two order parameters derived from different TAE models are also visualized, with particles color-coded according to their long-term propensities.

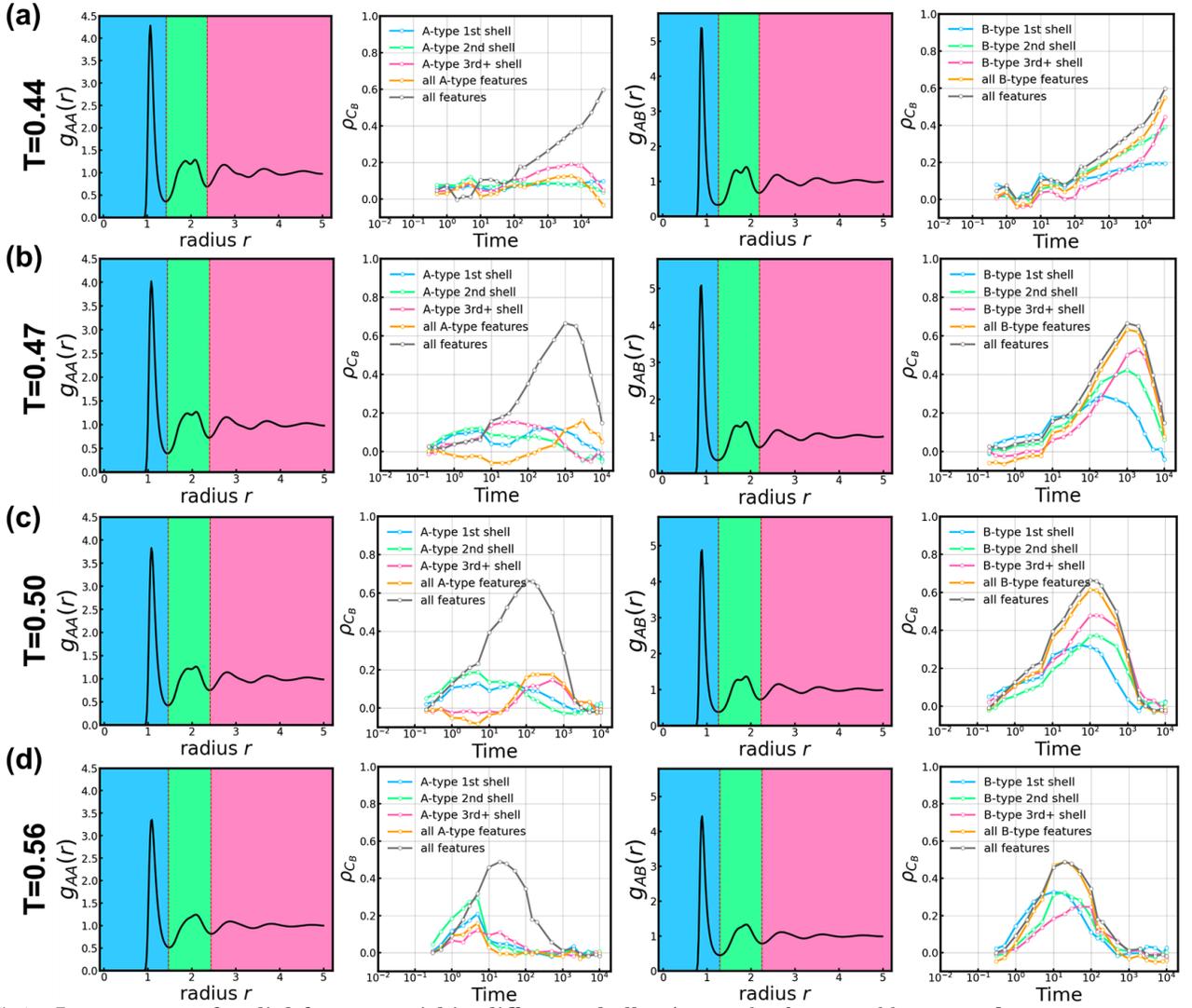

**FIG. 7: Importance of radial features within different shells.** A-particles from equilibrium configurations at temperatures (a) $T = 0.44$, (b) $T = 0.47$, (c) $T = 0.50$, and (d) $T = 0.56$ are embedded using 200-dimensional radial features. These features are then subsampled based on particle types (left two columns for A-particles, right two columns for B-particles) and shells of pair correlation functions (regions colored blue, green, and pink). Independent TAE models are constructed with different input features. For each model, the correlation between propensity and the order parameter $\lambda_1$ is calculated and visualized. The different curves for correlation are obtained using features from same colored regions in the pair correlation function.

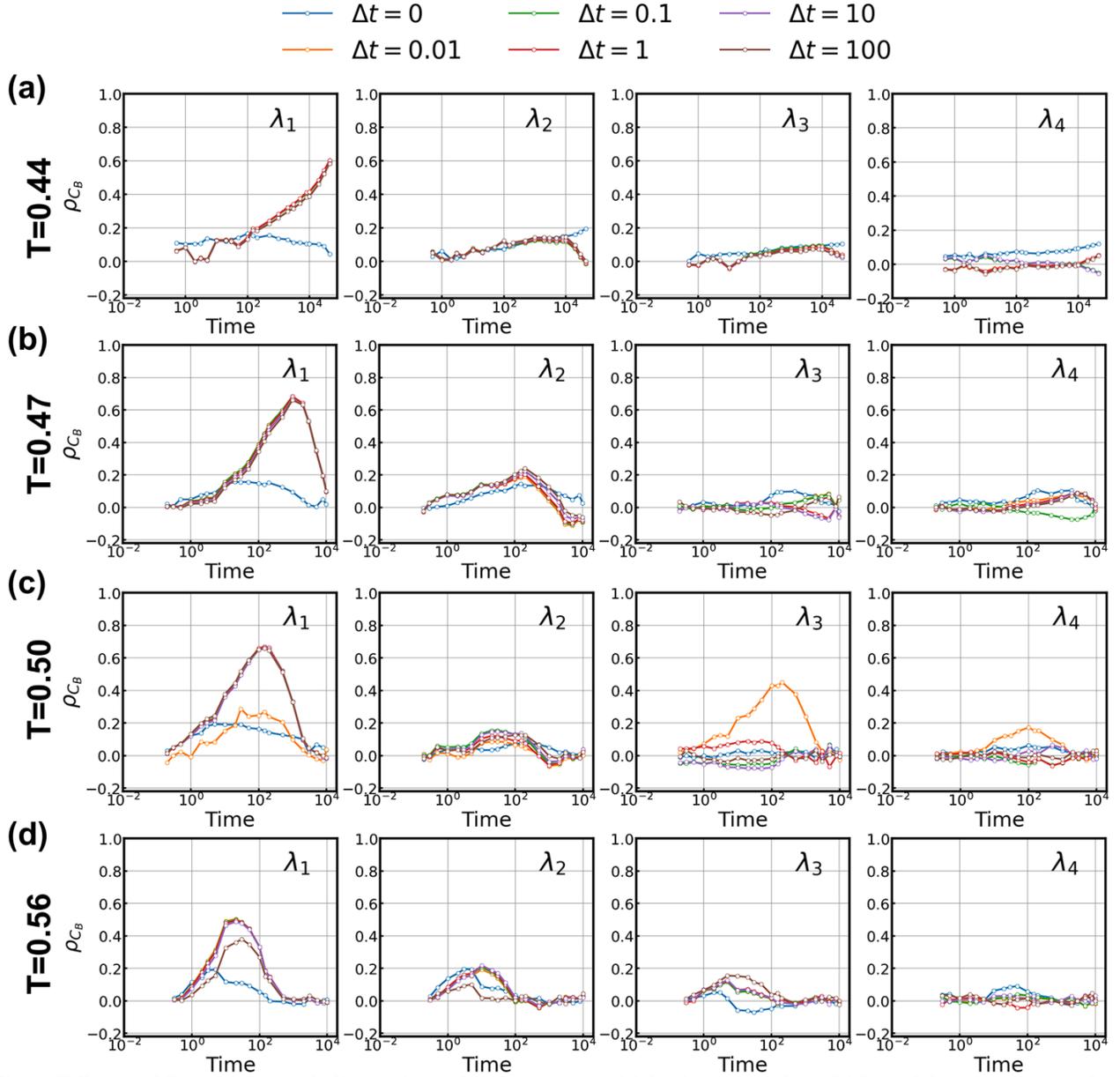

**FIG. 8: Effects of Lag Time and dimension of latent space.** Multiple independent TAE models are constructed with 4-dimensional latent space and trained with various lag times $\Delta t = 0, 0.01, 0.1, 1, 10, 100$ by utilizing radial features derived from time-lagged configurations at different temperatures: (a) T=0.44, (b) T=0.47, (c) T=0.50 and (d) T=0.56. The correlation between four order parameters and bond-breaking propensities are calculated and visualized, respectively.

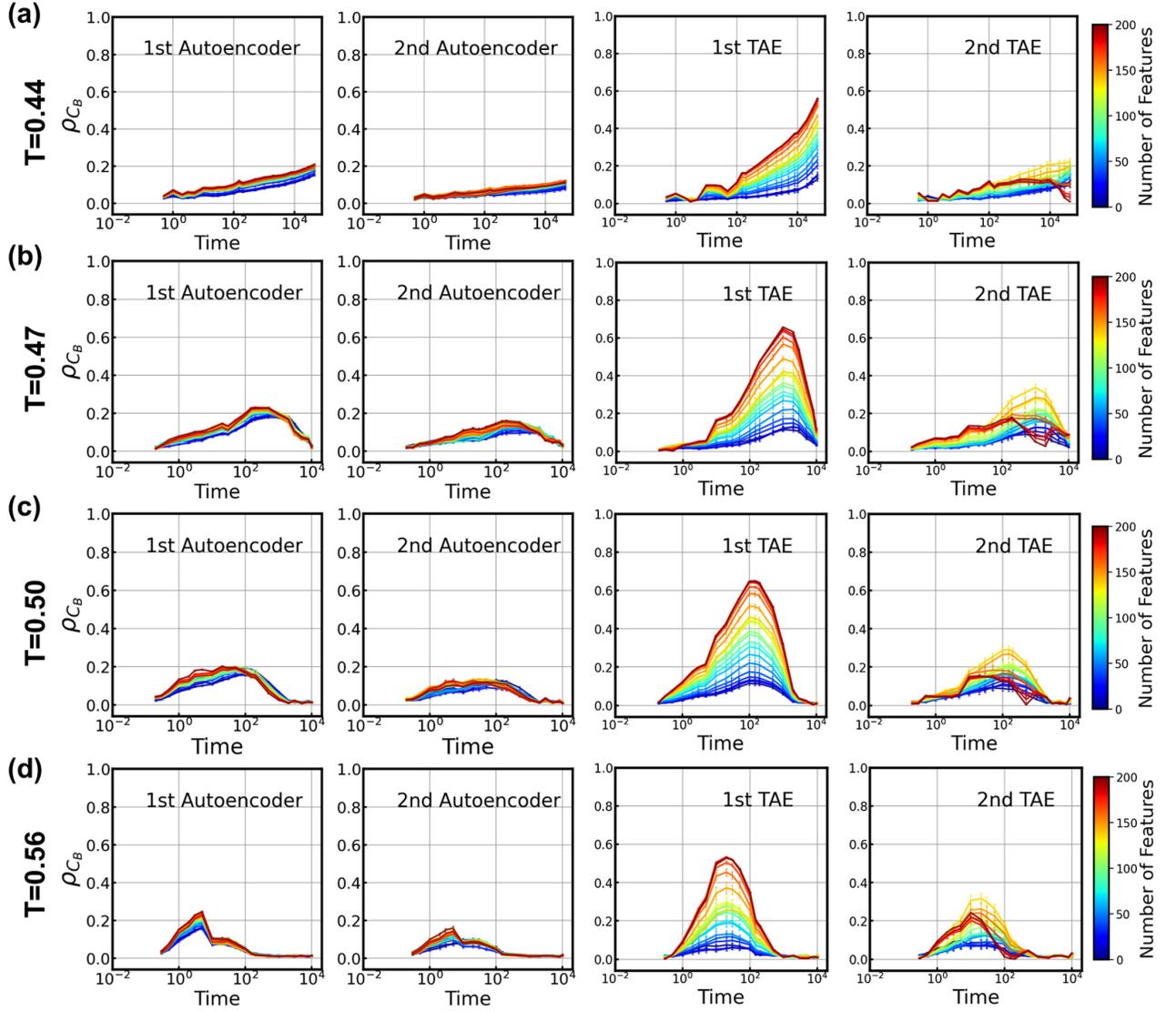

**FIG. 9: Impact of input feature dimension on the quality of order parameters.** The 200-dimensional radial density features are randomly subsampled into different dimensions and used to train both AE and TAE models at four respective temperatures: (a) $T = 0.44$, (b) $T = 0.47$, (c) $T = 0.50$, and (d) $T = 0.56$. The number of hidden neurons is consistently set as five times the input feature dimensions for all the models, and the lag time for TAE is fixed at $\Delta t = 0.1$. The subsamplings are repeated twenty times for a specific dimension of input features, and the error bars are estimated accordingly.



\end{document}